\documentclass[prb,aps,twocolumn,superscriptaddress,10pt,amsmath,amssymb,nofootinbib,floatfix,longbibliography]{revtex4-2}

\usepackage{cabin}
\usepackage{amsmath}
\usepackage{graphicx}
\usepackage{multirow}
\makeatletter
\begin{document}

\title{Piezoelectricity in hafnia}
\author{Sangita Dutta}
\affiliation{Materials Research and Technology Department,
             Luxembourg Institute of Science and Technology,
             5 avenue des Hauts-Fourneaux,
             L-4362 Esch/Alzette, Luxembourg}
\affiliation{Department of Physics and Materials
            Science, University of Luxembourg, 41 Rue du Brill, Belvaux L-4422,
            Luxembourg}
\author{Pratyush Buragohain}%
\affiliation{Department of Physics and Astronomy,
University of Nebraska-Lincoln,
Lincoln, NE 68588-0299, USA}
\author{Sebastjan Glinsek}%
\affiliation{Materials Research and Technology Department,
             Luxembourg Institute of Science and Technology,
             5 avenue des Hauts-Fourneaux,
             L-4362 Esch/Alzette, Luxembourg}%
\author{Claudia Richter}
\affiliation{NaMLab gGmbH,
Noethnitzer Strasse 64 a,
01187 Dresden, Germany}
\author{Hugo Aramberri}
\affiliation{Materials Research and Technology Department,
             Luxembourg Institute of Science and Technology,
             5 avenue des Hauts-Fourneaux,
             L-4362 Esch/Alzette, Luxembourg}
\author{Haidong Lu}
\affiliation{Department of Physics and Astronomy,
University of Nebraska-Lincoln,
Lincoln, NE 68588-0299, USA}
\author{Uwe Schroeder}
\affiliation{NaMLab gGmbH,
Noethnitzer Strasse 64 a,
01187 Dresden, Germany}
\author{Emmanuel Defay }
\affiliation{Materials Research and Technology Department,
             Luxembourg Institute of Science and Technology,
             5 avenue des Hauts-Fourneaux,
             L-4362 Esch/Alzette, Luxembourg}
\author{Alexei Gruverman }
\affiliation{Department of Physics and Astronomy,
University of Nebraska-Lincoln,
Lincoln, NE 68588-0299, USA}

\author{Jorge \'I\~niguez}
\affiliation{Materials Research and Technology Department,
             Luxembourg Institute of Science and Technology,
             5 avenue des Hauts-Fourneaux,
             L-4362 Esch/Alzette, Luxembourg}
\affiliation{Department of Physics and Materials
            Science, University of Luxembourg, 41 Rue du Brill, Belvaux L-4422,
            Luxembourg}

\begin{abstract}
Because of its compatibility with semiconductor-based technologies, hafnia (HfO$_{2}$) is today's most promising ferroelectric material for applications in electronics. Yet, knowledge on the ferroic and electromechanical response properties of this all-important compound is still lacking. Interestingly, HfO$_{2}$ has recently been predicted to display a negative longitudinal piezoelectric effect, which sets it apart form classic ferroelectrics (e.g., perovskite oxides like PbTiO$_{3}$) and is reminiscent of the behavior of some organic compounds. The present work corroborates this behavior, by first-principles calculations and an experimental investigation of HfO$_{2}$ thin films using piezoresponse force microscopy. Further, the simulations show how the chemical coordination of the active oxygen atoms is responsible for the negative longitudinal piezoelectric effect. Building on these insights, it is predicted that, by controlling the environment of such active oxygens (e.g., by means of an epitaxial strain), it is possible to change the sign of the piezoelectric response of the material.
\end{abstract}

\maketitle

Hafnia (HfO$_{2}$) is a well-known material in the electronics industry, since its introduction in 2007 by Intel as a convenient gate dielectric for field-effect transistors (FETs) \cite{bohr07}. The announcement of ferroelectricity in this compound a few years later \cite{boscke11,muller12} caused great excitement, as it opened the door to the development of (inexpensive, easy to process) electronic devices that could benefit from a switchable polarization, e.g., memories based on ferroelectric FETs. Ever since, a lot of efforts have focused on understanding and controlling ferroelectricity in HfO$_{2}$, taking advantage of the unique possibilities it may offer (e.g., ferroelectric negative-capacitance effects \cite{iniguez19}) and advancing towards commercial devices. By now, ferroelectric hafnia has gathered the interest of engineers, materials scientists and physicists alike, being one of today's best studied and most promising materials.

As compared to traditional soft-mode ferroelectrics (e.g., perovskite oxides like PbTiO$_{3}$ or BaTiO$_{3}$ \cite{lines-book1977}), hafnia displays many peculiar features that we are only starting to understand. For example, recent theoretical work suggests that hafnia's ferroelectricity is not proper in character \cite{delodovici21}, yet switchable, which sets this compound apart from all ferroelectrics used so far in applications. Further, the nature of its (anti)polar instabilities is such that very narrow domains, and very narrow domain walls, occur naturally in it \cite{lee20,noheda00}; in effect, this makes HfO$_{2}$ a quasi-2D ferroelectric, and may explain the resilience of its polar phase in nanometric samples. In fact, unlike traditional materials, HfO$_{2}$ seems to improve its ferroelectric properties as the samples decrease in size; in fact, the first reports of ferroelectricity in thick films or bulk samples are very recent \cite{schenk20,xu21}. In sum, from both applied and fundamental perspectives, ferroelectric hafnia is revealing itself as a very interesting compound.

Comparatively, the electromechanical response properties of ferroelectric HfO$_{2}$ have received little attention so far, although we believe this situation will quickly change. Indeed, the processing advantages that hafnia offers (as compared to perovskite oxides) make it a viable candidate for applications as a piezoelectric (e.g., in piezotronics, radio-frequency filters) where it might potentially compete with wurtzite compounds (e.g., AlN, ZnO).

At a more fundamental level, it has been recently predicted from first principles \cite{liu19b,liu20} that the usual ferroelectric phase of HfO$_{2}$ (orthorhombic with space group $Pca2_{1}$) presents a negative longitudinal piezoresponse, i.e., that compressing the material along the polarization direction will result in an enhancement of its polar distortion. If confirmed, this property would widen even more the gap between HfO$_{2}$ and the ferroelectric perovskite oxides, all of which behave in exactly the opposite way \cite{roedel09}. Intriguingly, though, existing experimental measurements of hafnia's piezoresponse \cite{schenk20} suggest a perovskite-like behavior (i.e., a positive longitudinal effect), and thus contradict the first-principles predictions. Further, we still lack a satisfying physical picture explaining the atomistic origin of the predicted negative piezoresponse, a simple understanding that would allow us to propose ways to control and optimize the effect. Hence, in our opinion, the piezoelectric response of hafnia is an open problem.

Here we present a first-principles and experimental investigation of the piezoelectric properties of HfO$_{2}$. First, we confirm the negative longitudinal effect, from both theory and experiment, notwithstanding the fact that the experimental result is sample dependent. Second, based on our first-principles simulations, we provide a simple and plausible explanation of the atomistic mechanisms controlling the effect. Further, based on this understanding, we predict that the ferroelectric phase of hafnia can be modified (by epitaxial strain) to either enhance or reduce the negative longitudinal piezoresponse, or even change its sign. We conclude with a brief discussion of the implications of our results, and an outlook of the challenges and opportunities ahead.

\vspace{5mm}{\bf\cabin Results and discussion}

In the following we present our simulation and experimental
results. In most occasions we discuss in parallel our findings for
HfO$_{2}$ and the corresponding results for a representative
ferroelectric perovskite, which allows us to better highlight the
specificity of hafnia as compared with classic materials. For the
perovskite, we consider PbTiO$_{3}$ at the theoretical level and
PbZr$_{1-x}$Ti$_{x}$O$_{3}$ (PZT, with $x = 0.6$) at the experimental
level. In our experimental presentation, we also show results for polyvinylidene fluoride (PVDF), a compound with a well-characterized negative longitudinal piezoresponse \cite{sharma11}.

\begin{figure}[t!]
    \centering
    \includegraphics[width=0.5\textwidth]{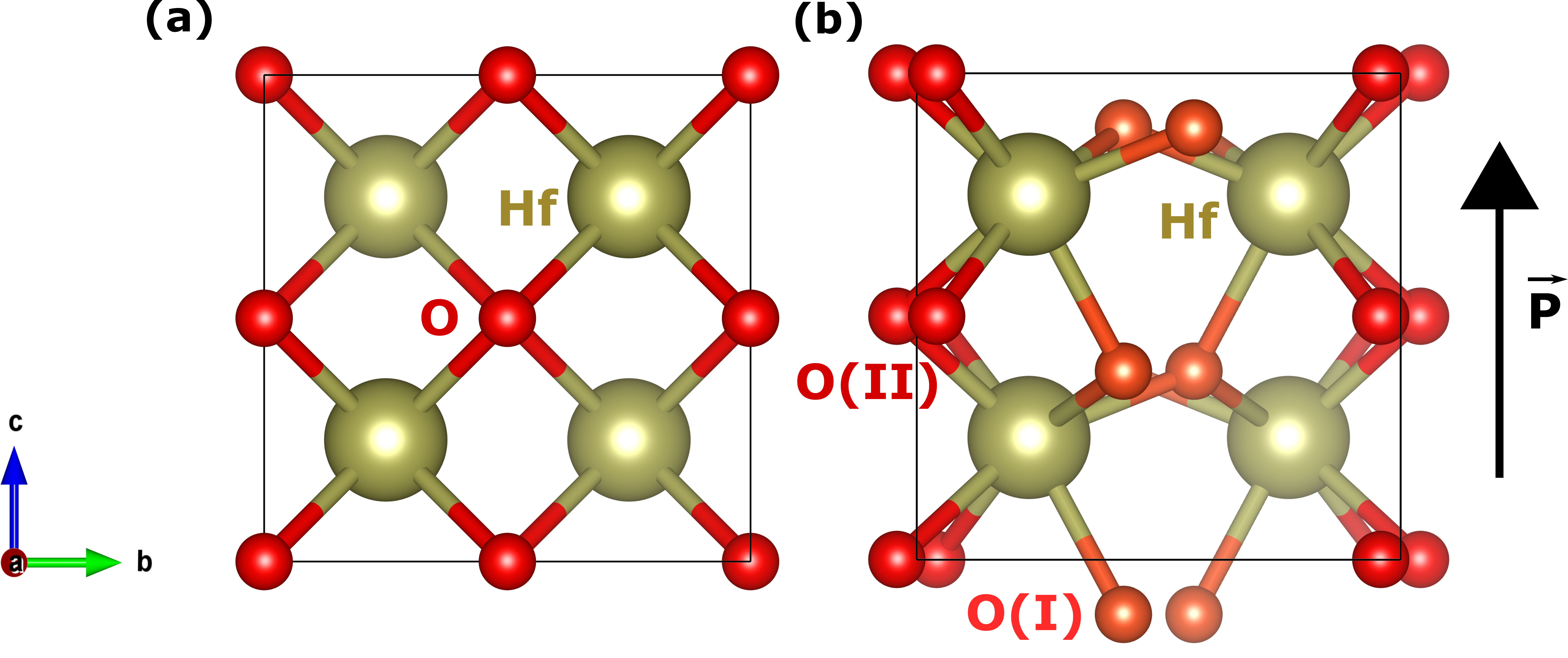}
    \caption{Structure of the cubic paraelectric (a) and orthorhombic ferroelectric (b) polymorphs of HfO$_{2}$. In the cubic $Fm\bar{3}m$ phase, all Hf and O atoms are equivalent by symmetry. In the ferroelectric $Pca2_{1}$ structure we have two symmetry-inequivalent sets of oxygen atoms -- labeled O(I) (shown in orange) and O(II) (red), respectively --, while all Hf atoms are equivalent. In panel~(b), the arrow indicates the positive spontaneous polarization of the structure shown, which is basically related to the vertical downward shift of the O(I) atoms from their high-symmetry position in the cubic phase.}
    \label{fig:hfo2-1}
\end{figure}

{\bf\cabin  Basic first-principles predictions.} We start our
first-principles investigation by relaxing the most usual
ferroelectric polymorph of HfO$_{2}$, with $Pca2_{1}$ space group. The
obtained solution (see Figure~\ref{fig:hfo2-1} and Table~S1 of the
Supporting Information) agrees well with previous results in the
literature \cite{huan14,dutta20}.

To fix ideas, noting that the sign of the piezoelectric coefficients depends on the sign of the spontaneous polarization \cite{strukov-book1998}, in the following we always work with the ferroelectric state with positive polarization along the third Cartesian direction ($c$ axis in Figure~\ref{fig:hfo2-1}). This polarization is essentially related to the downward displacement of the O(I) anions (orange atoms in Figure~\ref{fig:hfo2-1}) from their high-symmetry position in the cubic paraelectric reference; we find it to be $P_{3} = 54.75$~$\mu$C/cm$^{2}$, in agreement with previous literature \cite{huan14,dutta20}.

We now consider the piezoelectric tensor \cite{nye-book1985}
\begin{equation}
{e}_{\alpha j}=\frac{\partial P_{\alpha}}{\partial \eta_j} \; ,
\end{equation}
where $P_{\alpha}$ is the polarization component along Cartesian
direction $\alpha$ and $\eta_{j}$ is a symmetric strain labeled using
Voigt notation \cite{nye-book1985}. For analysis purposes, it is convenient to decompose the piezoelectric response into a frozen-ion contribution ($\bar{e}_{\alpha j}$) and a lattice-mediated part (defined as the difference $e_{\alpha j}-\bar{e}_{\alpha j}$). $\bar{e}_{\alpha j}$ is obtained by freezing the atoms in their unperturbed equilibrium
positions, so they cannot respond to the applied strain; it thus
captures a purely electronic effect. Finally, from knowledge of the
$e_{\alpha j}$ and the elastic constants $C_{jk}$, one can obtain 
\begin{equation}
{d}_{\alpha j}= \frac{\partial P_{\alpha}}{\partial \sigma_j} =
(C^{-1})_{jk} e_{\alpha k} = S_{jk} e_{\alpha k} \;
\end{equation}
where $\sigma_{j}$ is the $j$-th component of an applied external stress (in Voigt notation), ${\bf S}={\bf C}^{-1}$ is the elastic compliance, and we assume summation over repeated indices. Note that this ${\bf d}$ tensor is of interest, as it is the one most easily accessible in experiment and the one usually exploited in applications. Computing the piezoelectric tensors is straightforward using density functional perturbation theory (DFPT) \cite{wu05}.

\begin{table}[t!]
    \centering
    \begin{tabular}{lcrrr}
& Index & \multicolumn{1}{c}{ $\bar{e}$ } & \multicolumn{1}{c}{${e}$} & \multicolumn{1}{c}{${d}$} \\
\hline
VASP        & 31 & $-$0.37 & $-$1.31 & $-$1.71 \\
            & 32 & $-$0.34 & $-$1.33 & $-$1.77 \\
            & 33 &  0.62 & $-$1.44 & $-$2.51 \\
            & 15 & $-$0.28 & $-$0.20 & $-$2.03 \\
            & 24 & $-$0.20 &  0.64 &  6.74 \\ \hline
ABINIT      & 31 & $-$0.39 & $-$1.53 & $-$2.71 \\
            & 32 & $-$0.36 & $-$1.40 & $-$1.66 \\
            & 33 &  0.65 & $-$1.34 & $-$1.64 \\
            & 15 & $-$0.29 & $-$0.23 & $-$2.70 \\
            & 24 & $-$0.22 &  0.69 &  6.60 \\ \hline

\end{tabular}
    \caption{Calculated piezoelectric tensors for HfO$_{2}$. We show the total ($\mathbf{e}$) and frozen-ion ($\bar{\mathbf{e}}$) direct piezoelectric tensor (in C/m$^2$), as well as the total converse piezoelectric tensor $\mathbf{d}$ (in pm/V). Indices given in Voigt notation.}
    \label{tab:piezo-hfo2}
\end{table}

Table~\ref{tab:piezo-hfo2} shows the results we obtain for ${\bf e}$, $\bar{\bf e}$, and ${\bf d}$ using two different -- but essentially equivalent, both accurate -- implementations of Density Functional Theory (DFT). (We attribute the existing numerical differences mainly to the use of different pseudopotentials; see the Experimental Section.) We confirm a negative value of the $e_{33}$ coefficient, indicating that a positive strain (stretching) of the unit cell along the polar direction will yield a reduction of the
polarization. (Recall that we work with an unperturbed state with $P_{3}>0$ and $P_{1}=P_{2}=0$.) We further verify this result by performing a finite-difference calculation of the change in $P_{3}$ upon application of a small strain $\eta_{3}$. Our results are also in agreement with the DFT predictions previously published \cite{liu20}.

It is interesting to note that the lattice-mediated response is always larger than the frozen-ion response. In particular, Table~\ref{tab:piezo-hfo2} clearly shows that the lattice response is responsible for the negative value of $e_{33}$.

\begin{table}[t!]
    \centering
    \begin{tabular}{crrrr}
Index & \multicolumn{1}{c}{$\bar{C}$} & \multicolumn{1}{c}{${C}$} & \multicolumn{1}{c}{$\bar{S}$} & \multicolumn{1}{c}{${S}$} \\
\hline
 11 & 465.3 & 413.6 &  2.65 &  2.99 \\
 12 & 181.3 & 162.3 & $-$0.78 & $-$0.99 \\
 13 & 151.7 & 123.4 & $-$0.61 & $-$0.60 \\
 22 & 485.0 & 407.8 &  2.59 & 3.08  \\
 23 & 165.6 & 132.8 & $-$0.69 & $-$0.73 \\
 33 & 445.7 & 394.6 &  2.71 &  2.97 \\
 44 & 127.0 &  94.4 &  7.87 &  10.59 \\
 55 & 116.7 &  98.0 &  8.56 &  10.20 \\
 66 & 169.3 & 140.4 &  5.90 &  7.12 \\ \hline
\end{tabular}
    \caption{Computed elastic ($\mathbf{C}$, in GPa) and compliance ($\mathbf{S}$, in TPa$^{-1}$) tensors of the ferroelectric phase of HfO$_{2}$. We show the total and frozen-ion (barred) effects. Indices in Voigt notation.}
    \label{tab:elastic-hfo2}
\end{table}

Table~\ref{tab:elastic-hfo2} shows the results obtained for the elastic and compliance tensors, which allow us to compute the ${\bf d}$ tensor in Table~\ref{tab:piezo-hfo2}. While the relationship between ${\bf e}$ and ${\bf d}$ is not trivial in materials with a relatively low symmetry (as is the case of ferroelectric HfO$_{2}$), we do obtain a negative $d_{33} = -2.51$~pm/V. Hence, we predict that, upon application of a compressive stress $\sigma_{3} < 0$, $P_{3}$ will increase.

\begin{table}[t!]
    \centering
\begin{tabular}{crrr}
Index & \multicolumn{1}{c}{ $\bar{e}$ } & \multicolumn{1}{c}{${e}$} & \multicolumn{1}{c}{${d}$} \\
\hline
  31 &  0.23 &  1.62 & $-$39   \\
  33 & $-$0.30 &  4.95 &  208  \\
  15 &  0.03 &  4.58 &  78   \\ \hline
            \end{tabular}
    \caption{Same as Table~\ref{tab:piezo-hfo2}, but for the ferroelectric phase of PbTiO$_{3}$.}
    \label{tab:piezo-pto}
\end{table}

Table~\ref{tab:piezo-pto} shows results for the piezoelectric tensors of the ferroelectric phase of PbTiO$_{3}$ (tetragonal, with space group $P4mm$). We find that the $e_{\alpha j}$ coefficients are generally larger for the perovskite than for HfO$_{2}$ (by a factor of up to 4, if we focus on their absolute values). Interestingly, the difference becomes much greater for the $d_{\alpha j}$ coefficients, as PbTiO$_{3}$ presents values between 1 and 2 orders of magnitude larger than those of hafnia: for example, we get a $d_{33}$ of 208~pm/V for PbTiO$_{3}$ and $-$2.51~pm/V for HfO$_{2}$. The obtained giant $d_{33}$ response of lead titanate agrees with previous experimental \cite{haun87} and theoretical \cite{Kvasov2016} reports. Interestingly, our calculations indicate that the main reason behind this result is the softness of the compound along the polarization direction: we get $C_{33} = 51.8$~GPa and $S_{33} = 48.72 $~TPa$^{-1}$ for PbTiO$_{3}$ (the full tensors are given in Table~S2 of the Supporting Information), which contrast with the much stiffer case of HfO$_{2}$ ($S_{33}= 2.97$~TPa$^{-1}$, see Table~\ref{tab:elastic-hfo2}).

When considering the experimental manifestation of the negative $d_{33}$ predicted for hafnia, one has to take into account an important feature of most samples: they are polycrystalline. Hence, typically, the measured piezoelectric response will not correspond to a unique well-defined orientation, but to an effective average. If we assume a sample composed of randomly-oriented grains, and poled so that all grains display a polarization with a positive $P_{3}$ component, we can estimate an effective $d_{33, {\rm eff}}$ as \cite{damjanovic06,heywang-book2008}
\begin{widetext}
\begin{equation}
\begin{split}
d_{33, {\rm eff}} & = \left\langle \cos\theta
\left[(d_{15}+d_{31})(\sin^2{\theta} \sin^2{\varphi}) +
  (d_{24}+d_{32})(\sin^2{\theta}\cos^2{\varphi})+d_{33}\cos^2{\theta}
  \right] \right\rangle  \\
  & =
  (d_{15}+d_{31})\langle\cos{\theta}\sin^2{\theta}\rangle\langle \sin^2{\varphi}\rangle +
  (d_{24}+d_{32})\langle\cos{\theta}\sin^2{\theta}\rangle\langle\cos^2{\varphi}\rangle+d_{33}\langle\cos^3{\theta}\rangle \\
  & = \frac{1}{3\pi}(d_{15}+d_{31}+d_{24}+d_{32})+\frac{4}{3\pi}d_{33}\; ,
\end{split}
\end{equation}
\end{widetext}
where $\langle ... \rangle$ indicates an average over the Euler angles $\varphi$ and $\theta$, which span all possible orientations with $P_{3}>0$ (i.e., $0<\theta<\pi/2$ and $0<\varphi<2\pi$). By calculating this average we obtain $d_{33,{\rm eff}} = -0.94$~pm/V, suggesting that even in polycrystalline HfO$_{2}$ samples we expect to measure a negative longitudinal piezoresponse.

{\bf\cabin Experimental confirmation.} To determine experimentally the sign of the effective piezoelectric coefficient $d_{33,{\rm eff}}$ in hafnia, we carry out comparative dynamic piezoelectric measurements by means of piezoresponse force microscopy (PFM), using materials with known piezoelectric coefficients as a reference. In PFM, application of an alternating (ac) electric field to the sample via a conductive tip results in an oscillation with the frequency of the applied field, due to the converse piezoelectric effect \cite{Gruverman2019,Hong2021}. The amplitude and phase of the oscillation provide information about the magnitude and sign of $d_{33,{\rm eff}}$, respectively (see Section~S1 and Figures~S1 to S3 of the Supporting Information).

\begin{figure}[t!]
    \centering
    \includegraphics[width=0.48\textwidth]{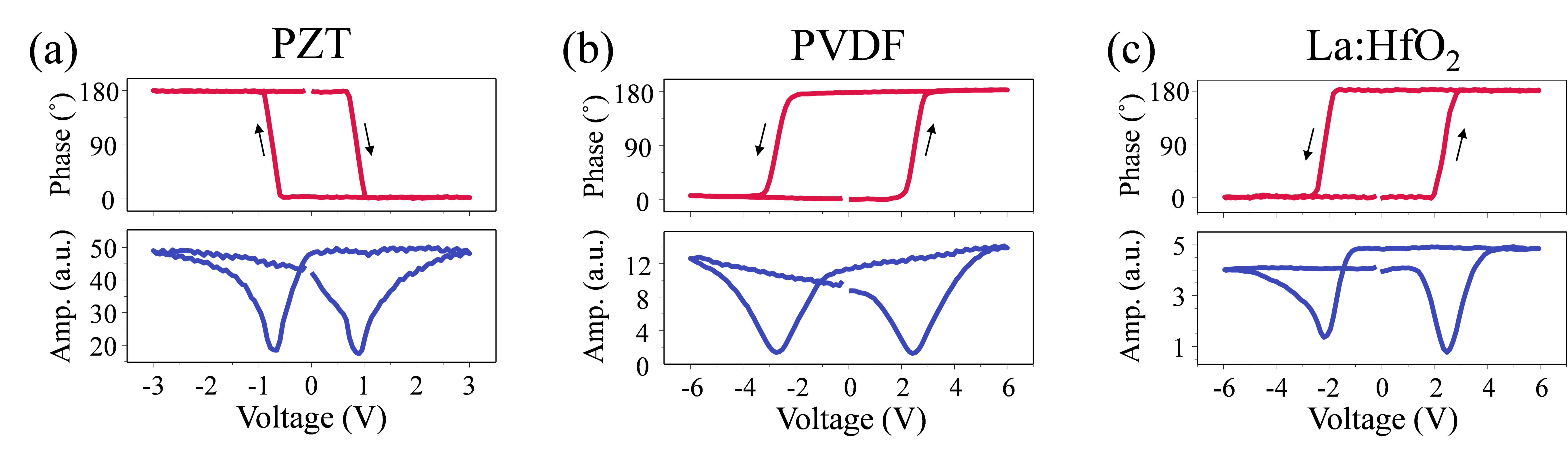}
    \caption{Dynamic measurements of piezoelectricity using
      piezoresponse force microscopy (PFM). (a-c) PFM phase (top
      panel) and amplitude (bottom panel) loops measured in the
      IrO$_{2}$/PZT/Pt capacitor (a), PVDF film (b), and
      Ti/Pt/TiN/La:HfO$_{2}$/TiN capacitor (c). The loops were
      obtained in the bias-off mode to minimize the electrostatic
      contribution to the PFM signal.}
    \label{fig:exp}
\end{figure}

In a material with a positive longitudinal piezoresponse, such as PZT \cite{Kholkin2001}, the sample oscillation will be in phase with the driving electric field when the polarization is oriented downward, while it will be in anti-phase when the polarization is oriented upward. Additionally, in the switching spectroscopy mode of PFM, a pulsed low-frequency triangular waveform is superimposed on the ac waveform to generate a piezoelectric strain hysteresis loop related to the local polarization reversal \cite{Jesse2006}. Thus, the sense of rotation of the PFM phase loops in Figure~\ref{fig:exp} is directly related to the sign of the $d_{33,{\rm eff}}$.
For example, a clockwise rotation of the PFM phase signal is indicative of a positive $d_{33,{\rm eff}}$ coefficient, as is illustrated by the PFM phase hysteresis loop measured in the IrO$_{2}$/PZT/Pt capacitor in Figure~\ref{fig:exp}(a). In this case, the phase signal is in phase (anti-phase) with the ac modulation field at the far positive (negative) dc bias, which generates the downward (upward) orientation of the polarization. In contrast, in a material with a negative $d_{33,{\rm eff}}$, such as PVDF \cite{1990JaJAP..29..675F,Katsouras2016}, the phase signal is in phase (anti-phase) with the ac modulation field at the far negative (positive) dc bias corresponding to the upwards (downwards) orientation of the polarization, resulting in anti-clockwise rotation in the PFM phase loop (Figure~\ref{fig:exp}(b)). (See the Experimental Section for details on our PZT and PVDF samples.)
Under the same conditions, as shown in Figure~\ref{fig:exp}(c), the PFM phase loop measured in TiN/La:HfO$_{2}$/TiN capacitors (with 20~nm-thick La:HfO$_{2}$) exhibits an anti-clockwise rotation similar to that of PVDF films and opposite to that of PZT capacitors. (See the Experimental Section for details on our HfO$_{2}$ samples; note that similar results were obtained for 10~nm-thick samples.) Clearly, this behavior is indicative of the negative sign of $d_{33,{\rm eff}}$ in the La:HfO$_{2}$ film. We estimate $d_{33,{\rm eff}}$ to be between $-$2~pm/V and $-$5~pm/V (Figure~S2 (c)), in excellent agreement with our theoretical result between $-$0.9~pm/V (in the polycrystalline disordered limit) and $-$2.5~pm/V (in the single-crystal limit).

It is important to note here that evidence for a positive longitudinal piezoresponse $d_{33,{\rm eff}}$ in hafnia has been reported in the experimental literature, for example, for ultrathin (10~nm) Si-doped HfO$_{2}$ films \cite{muller11c}, for thicker (70~nm) Y-doped films \cite{Starschich2014}, or for La-doped films with thicknesses up to 1~$\mu$m \cite{schenk20}. Indeed, in the course of this work, we found ourselves that the application of the same experimental protocol to other HfO$_{2}$ films (thicker, grown by different means) yields a positive longitudinal effect. Hence, one may wonder what is especial about the La:HfO$_{2}$ films for which we get a negative effect. May it be extrinsic?

Interestingly, the La:HfO$_{2}$ films studied here were characterized in Ref.~\cite{schenk19} using grazing-incidence X-ray diffraction. In that work, it was determined that these La:HfO$_{2}$ films present a strong out-of-plane texture, leading to relatively high remnant polarization compared to other dopants. This stronger texture may cause local strains in the films, and potentially act as an extrinsic factor affecting the local electromechanical properties. However, since all the domains measured in our study showed a negative response, we tend to believe that the effect observed in this work is intrinsic.

As far as we can tell, there is only one other publication suggesting a similar negative longitudinal piezoresponse. In 2019 Chouprik {\sl et al}. \cite{chouprik19} reported ``anomalous'' switching in 10~nm-thick films of Hf$_{0.5}$Zr$_{0.5}$O$_{2}$, which in PFM appeared as a polarization reversal against the applied electric field, and could thus be interpreted as a negative piezoresponse. This behavior was observed in about 20~\% of the domains of the pristine films. However, after ac field cycling, all domains showed a positive piezoresponse, which was attributed to the release of trapped charges, a signature of an extrinsic effect.

Hence, there is experimental evidence that the piezoelectric properties of HfO$_{2}$ are sample and sample-history dependent. This suggests that a careful and systematic characterization will be needed to determine the factors (intrinsic or extrinsic) controlling the piezoresponse, including its sign. 

{\bf\cabin Origin of the negative piezoresponse.} Our DFPT calculations allow us to track down the computed negative value of
$e_{33}$. As presented in the Section~S2 of the Supporting Information, within perturbation theory \cite{wu05} we write the piezoelectric tensor as
\begin{equation}
    e_{\alpha j}=\bar{e}_{\alpha
      j}+\Omega_{0}^{-1}Z_{m\alpha}(\Phi^{-1})_{mn}\Lambda_{nj} \; ,
\end{equation}
where the second term on the right-hand side of this equation shows
that the lattice-mediated part of ${\bf e}$ depends on the unit-cell
volume of the unperturbed system ($\Omega_{0}$), the Born effective
charge tensor (${\bf Z}$, which quantifies the polarization change
caused by atomic displacements), the force-constant matrix
($\boldsymbol{\Phi}$, i.e., the second derivatives of the energy with
respect to atomic displacements) and the force-response internal
strain tensor ($\boldsymbol{\Lambda}$, which quantifies the atomic forces that appear when a strain is applied). Here, $m$ is a combined index that runs over all atoms in the unit cell and the three spatial directions.

By inspecting the calculated tensors for HfO$_{2}$, and by comparing
with those obtained for PbTiO$_{3}$, we can identify the atomistic
underpinnings of the sign of $e_{33}$.

First off, let us note that there is nothing peculiar concerning the
force-constant matrices $\boldsymbol{\Phi}$: for both materials, these
matrices reflect the fact that the ferroelectric phase is a stable
equilibrium state. Hence, they are positively defined tensors without
any feature that is relevant to the present discussion.

As for the Born effective charges $\boldsymbol{Z}$ (Table~S3
in the Supporting Information), they have the expected signs and are
relatively large in magnitude: we get values over $+5$ for Hf
and below $-2.5$ for O, exceeding the nominal respective charges of
$+4$ and $-2$. This feature reflects a mixed ionic-covalent character
of the chemical bonds in the material, and is typical of other
ferroelectrics like e.g. PbTiO$_{3}$ itself (see the Born charges we
obtain for PbTiO$_{3}$ in Table~S4 of the Supporting
Information). Additionally, because of the relatively low site
symmetries in the ferroelectric phase of HfO$_{2}$, in this compound the
charge tensors present small non-zero off-diagonal components. While
this feature does set HfO$_{2}$ apart from PbTiO$_{3}$, we checked it
has no influence in the sign of $e_{33}$. In conclusion, the ${\bf Z}$
tensors do not explain the differentiated behavior of $e_{33}$ in these two
compounds.

\begin{table}[t!]
    \centering
    \begin{tabular}{llccccccr}
   \multirow{3}{*}{Hf}   &\multirow{3}{*}{\Bigg[}& $-$4.02  & \phantom{-} 0.00 & $-$2.29      & \phantom{-} 8.39 & \phantom{-} 0.83 & \phantom{-} 0.79&\multirow{3}{*}{\Bigg]}\\
            &                      &                     $-$1.04  &\phantom{-} 11.52 & $-$1.40      & \phantom{-} 4.50 &\phantom{-} 3.58 &\phantom{-} 5.87& \\
            &                      &                     $-$1.87  & \phantom{-} 1.45 & $-$\textbf{4.07} &$-$2.83 &$-$2.61 & \phantom{-} 3.42& \\
    \multirow{3}{*}{O(I)} &\multirow{3}{*}{\Bigg[}&\phantom{-} 2.06  & $-$0.63 &\phantom{-}  1.09      &$-$5.87 &$-$3.29 & \phantom{-} 0.91 &\multirow{3}{*}{\Bigg]}\\
            &                      &                     $-$1.97  & $-$1.75 & $-$1.43      &$-$2.77 &$-$4.60 &$-$2.34& \\
            &                      &                     $-$1.61  &\phantom{-}  0.42 & \phantom{-} \textbf{3.22}  &\phantom{-} 1.36 & \phantom{-} 0.29 &$-$4.53& \\
    \multirow{3}{*}{O(II)}&\multirow{3}{*}{\Bigg[}&$-$5.87  & \phantom{-} 1.10 & $-$1.07      &\phantom{-} 4.67 &\phantom{-} 2.46 &\phantom{-} 0.06&\multirow{3}{*}{\Bigg]}\\
            &                      &                      \phantom{-} 0.79  & \phantom{-} 7.87 & \phantom{-} 0.54      &$-$1.71 &\phantom{-} 4.10 & \phantom{-} 0.52& \\
            &                      &                      \phantom{-} 3.46  & \phantom{-} 1.86 & \phantom{-} 0.85      & \phantom{-} 1.27 &$-$1.36 &\phantom{-} 5.44&
\end{tabular}
    \caption{Computed $\boldsymbol{\Lambda}$ tensors for the symmetry-inequivalent atoms of the ferroelectric phase of HfO$_{2}$ (in eV/\AA). The 3 rows correspond, respectively, to the 3 spatial directions; the 6 columns correspond, respectively, to the 6 strain indices in Voigt notation.}
    \label{tab:lambda-hfo2}
\end{table}

\begin{table}[t!]
    \centering
    \begin{tabular}{llccccccr}
    
\multirow{3}{*}{Pb}   &\multirow{3}{*}{\Bigg[}& 0     &   0   &  0         & 0    &\phantom{-} 3.34 &  0  &\multirow{3}{*}{\Bigg]}\\
            &                      &                       0     &   0   &  0         &\phantom{-} 3.15 & 0    &  0  & \\
            &                      &                      \phantom{-} 6.89   & \phantom{-} 6.89  &\phantom{-} 5.01        & 0    & 0    &  0  & \\
\multirow{3}{*}{Ti}   &\multirow{3}{*}{\Bigg[}& 0     &   0   &  0         & 0    &$-$0.51 &  0  &\multirow{3}{*}{\Bigg]}\\
            &                      &                      0     &   0   &  0         &\phantom{-}  0.50 & 0    &  0  & \\
            &                      &                     $-$3.18  &$-$3.18  &\phantom{-} 30.76       & 0    & 0    &  0  & \\
\multirow{3}{*}{O(I)} &\multirow{3}{*}{\Bigg[}& 0     &   0   &  0         & 0    &$-$0.44 &  0  &\multirow{3}{*}{\Bigg]}\\
            &                      &                     0     &   0   &  0         &$-$0.69 & 0    &  0  & \\
            &                      &                     $-$5.19  &$-$5.19  &$-$34.26      & 0    & 0    &  0  & \\
\multirow{3}{*}{O(II)}&\multirow{3}{*}{\Bigg[}& 0     &   0   &  0         & 0    &$-$2.94 &  0  &\multirow{3}{*}{\Bigg]}\\
            &                      &                        0     &   0   &  0         &$-$0.12 & 0    &  0  & \\ 
            &                      &                       $-$2.04  &\phantom{-} 3.55  &$-$0.81       & 0    & 0    &  0  &
\end{tabular}
    \caption{Same as Table~\ref{tab:lambda-hfo2} but for the ferroelectric phase of PbTiO$_{3}$.}
    \label{tab:lambda-pto}
\end{table}

Hence, we are left with the $\boldsymbol{\Lambda}$ tensors, which are
given in Tables~\ref{tab:lambda-hfo2} and \ref{tab:lambda-pto} for
HfO$_{2}$ and PbTiO$_{3}$, respectively. Let us focus on the ``33''
entries for each of the atom-specific tensors, i.e., the numbers quantifying
the atomic force along direction 3 (parallel to the polarization) caused by a positive strain $\eta_{3} > 0$ (stretching). In the case of PbTiO$_{3}$, the
strain-induced forces are positive for the cations (Pb and Ti) and
negative for the two symmetry-inequivalent oxygens in the unit
cell. This means that, in response to the vertical stretching of the
cell, the cations will tend to move up while the oxygens will tend to
move down. Since the unperturbed state has $P_{3}>0$, this movement
will clearly yield an increase of the polarization; hence, we have
$e_{33}>0$.

The situation is just opposite for HfO$_{2}$: in this case, a
stretching of the cell ($\eta_{3}>0)$ causes the Hf cations to move
down and the oxygens (particularly those of type~I) to move up. Since the
starting point has $P_{3}>0$, and since the Born charges in HfO$_{2}$
have the natural signs for cations and anions, these strain-induced
displacements will yield a reduction of the magnitude of the
polarization. This is indeed reflected in our computed $e_{33}<0$; we have
thus identified the atomistic origin of the effect.

{\bf\cabin Physical insight.} While the above discussion is clear from
a numerical point of view, it hardly provides us with a satisfying
physical understanding. Can we rationalize the mechanisms controlling
the sign of the strain-induced forces and, thus, of $e_{33}$?

In perovskite oxides, it is known that the structural instabilities of
the parent cubic phase (as, e.g., those leading to ferroelectricity) are
largely determined by steric and ion-size aspects, usually discussed
in terms of simple descriptors such as the Goldschmidt tolerance
factor \cite{goldschmidt26}. Ultimately, these effects are a reflection of the ions' tendency to optimize the chemical bonds in their first (nearest neighbor) coordination shell, as successfully captured by
phenomenological theories such as, for example, the bond-valence model \cite{brown09}.

\begin{figure}[t!]
    \centering
    \includegraphics[width=0.5\textwidth]{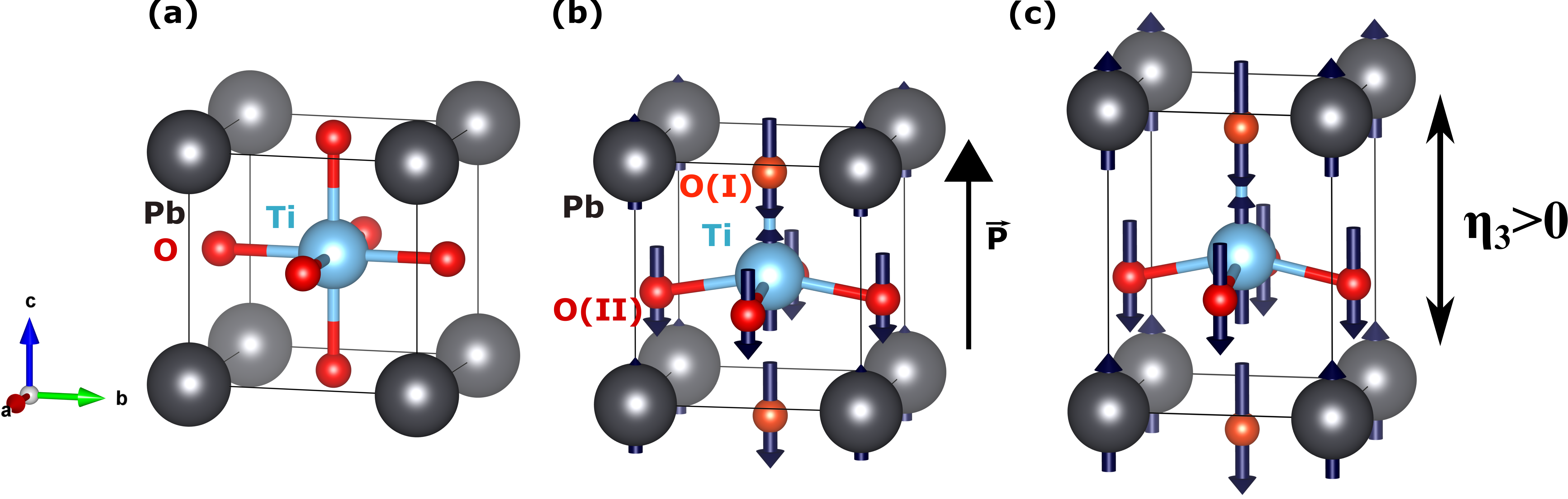}
    \caption{Cubic $Pm\bar{3}m$ paraelectric (a) and tetragonal $P4mm$
      ferroelectric (b) phases of PbTiO$_{3}$. The tetragonal phase
      presents two symmetry-inequivalent oxygen anions, colored
      differently and labeled by O(I) and O(II), respectively. In
      panel~(b) the arrow on the right marks the spontaneous
      polarization, which is essentially related to the upward
      displacement of the Pb and Ti cations with respect to the oxygen
      atoms (the arrows on the atoms mark such
      displacements). Panel~(c) is a sketch of the tetragonal phase
      subject to a tensile $\eta_{3} > 0$ strain (the strain is
      exaggerated for clarity); the arrows on the atoms indicate how
      they react in response to the strain.}
    \label{fig:pto}
\end{figure}

These bonding considerations readily allow us to understand the piezoelectric
response in PbTiO$_{3}$, a simple model case. In this compound, the
cubic paraelectric phase presents Ti and Pb cations that are
equidistant to 6 and 12 first-neighboring oxygens, respectively. Then,
as shown in Figure~\ref{fig:pto} (for a state with $P_{3}>0$), the
ferroelectric distortion results in a tetragonal structure where the
cations reduce the number of closest oxygens neighbors. For the sake of simplicity, let us focus on the case of the central Ti cation, which passes from being 6-fold coordinated in the paraelectric phase (panel~(a)) to having only 5 close oxygens in the ferroelectric state (panel~(b)). In fact, among these oxygens, there is one (the apical type~I oxygen that lies above Ti) forming the shortest (and strongest) Ti--O bond. (We know the details of this bond from previous theoretical works on PbTiO$_{3}$, which also show that the type~II oxygens are mainly bonded to the Pb cations \cite{walsh11}.) Imagine we now stretch the cell along the polarization direction ($\eta_{3}>0$), and assume that the atoms will rearrange in order to maintain the preferred length of the strongest bonds. For that to happen, as sketched in Figure~\ref{fig:pto}(c), the central Ti should move up and the mentioned O(I) oxygen that bonds with it should move down. This expectation is in perfect correspondence with our computed $\boldsymbol{\Lambda}$ tensor (see Table~\ref{tab:lambda-pto}). A similar argument applies to the displacements of the Pb and O(II) ions in reaction to $\eta_{3}>0$. Thus, this simple picture explains the positive $e_{33}$ obtained for PbTiO$_{3}$.
The situation in HfO$_{2}$ is harder to analyze, for two main reasons. First, the atomic chemical environments are far more complex than
in PbTiO$_{3}$ and identifying {\em dominant bonds} is not
trivial. (Our attempts at a clear-cut quantification -- e.g., by
inspecting the magnitude of the interatomic force constants -- were not convincing enough.) Second, the connection with a reference
high-symmetry structure is far more complicated than in PbTiO$_{3}$;
in fact, the largest distortions connecting the ferroelectric and
possible reference states (cubic $Fm\bar{3}m$, tetragonal $P4_{2}/nmc$) are {\em not} polar. (See Ref.~\cite{delodovici21} for details.)

\begin{figure}[t!]
    \centering
    \includegraphics[width=0.4\textwidth]{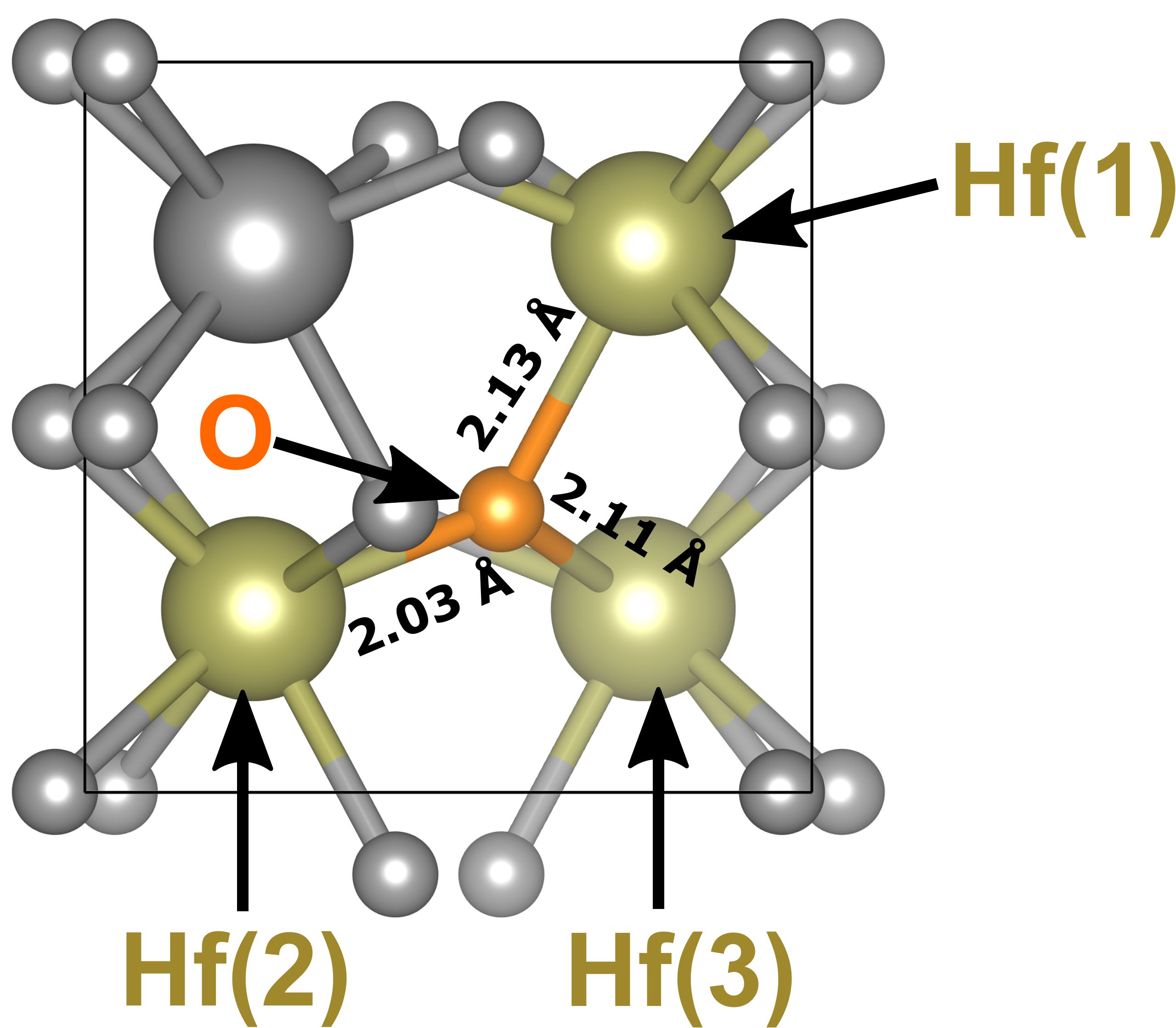}
    \caption{Sketch of the $Pca2_{1}$ ferroelectric phase of HfO$_{2}$ where we have highlighted a representative type~I oxygen as well as its three nearest-neighboring Hf cations. The reaction of type~I oxygens to strain largely determines the $e_{33}$ piezoresponse of the material.}
    \label{fig:hfo2-2}
\end{figure}

\begin{figure}[t!]
    \centering
    \includegraphics[width=0.5\textwidth]{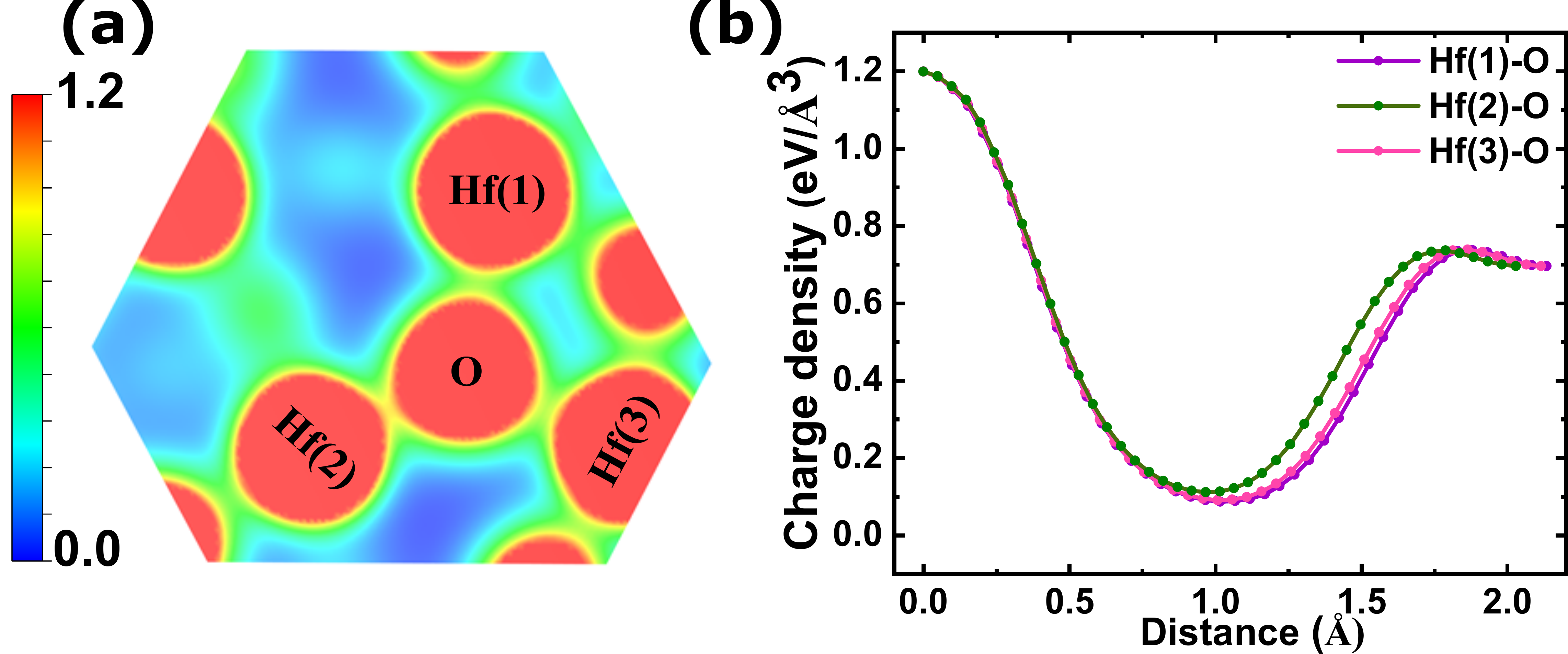}
    \caption{Computed electronic charge density for the unperturbed ferroelectric phase of HfO$_{2}$. Panel~(a) shows a contour plot of the charge density within a plane that approximately contains the oxygen atom highlighted in Figure~\ref{fig:pto} as well as its three nearest-neighboring Hf atoms. Panel~(b) shows the charge density along lines connecting the highlighted oxygen with each of its three nearest-neighboring Hf cations.}
    \label{fig:rho-eq}
\end{figure}

Nevertheless, if we try to explain the piezoresponse results for HfO$_{2}$ in simple bonding terms, we reach some useful conclusions. To fix ideas, let us
focus on the response to strain of the oxygen atom highlighted in
Figure~\ref{fig:hfo2-2}, which is a type~I oxygen in Table~\ref{tab:lambda-hfo2}. As indicated in Figure~\ref{fig:hfo2-2}, this oxygen has three nearest-neighboring Hf cations, the computed interatomic distances at equilibrium being 2.13~\AA\ for Hf(1), 2.03~\AA\ for Hf(2) and 2.11~\AA\ for Hf(3). Additionally, the computed equilibrium charge density in Figure~\ref{fig:rho-eq}(a) suggests that this oxygen forms similarly strong bonds with its three neighboring Hf cations.

As we have seen when discussing the $\boldsymbol{\Lambda}$ tensor
(Table~\ref{tab:lambda-hfo2}), when we stretch the cell along the
vertical direction ($\eta_{3}>0$), our central type~I oxygen will move up, while all the Hf cations will move down. (Note that all the hafnium atoms are equivalent by symmetry, hence they react in the same way to strain; see Table~S1 in the Supporting Information.) Therefore, we find that the Hf(1)--O bond seems to dominate the response, because the strain-driven atomic rearrangements contribute to preserve
the equilibrium length of this particular pair. This makes good
physical sense: since the three Hf--O links seem similarly strong,
the lattice response to the applied strain $\eta_{3}>0$ should be
dominated by the bond whose length changes the most due to the
perturbation; this is clearly the case of Hf(1)--O, which lies
almost parallel to the direction of the applied strain (i.e., the vertical in Figure~\ref{fig:hfo2-2}). Hence, this interpretation suggests that it is the peculiar atomic environment of the Hf and O ions in HfO$_{2}$ -- in particular, the specific Hf--O bond that lies nearly parallel to the polarization direction -- what determines the negative $e_{33}$ response.

While appealing, this qualitative picture may seem too
speculative. Nevertheless, it suggests that, by controlling the
chemical environment of the type~I oxygens that dominate the $e_{33}$ response, we should be able to affect the magnitude of the effect in a very
definite way. More precisely: by decreasing the Hf(1)--O distance, we
should be able to make this bond even more dominant and, thus, make $e_{33}$
more negative; conversely, by weakening the Hf(1)--O link, we should
have a response increasingly controlled by the Hf(2)--O and Hf(3)--O
pairs, which should result in a less negative $e_{33}$. We put this
hypothesis to a test in the next section.

{\bf\cabin Prediction of a tunable piezoresponse.} To control the bonds of interest and monitor their effect on $e_{33}$,
we simulate the $Pca2_{1}$ ferroelectric phase of HfO$_{2}$ subject to an isotropic epitaxial strain in the plane perpendicular to the polarization.

\begin{figure}[t!]
    \centering
    \includegraphics[width=0.5\textwidth]{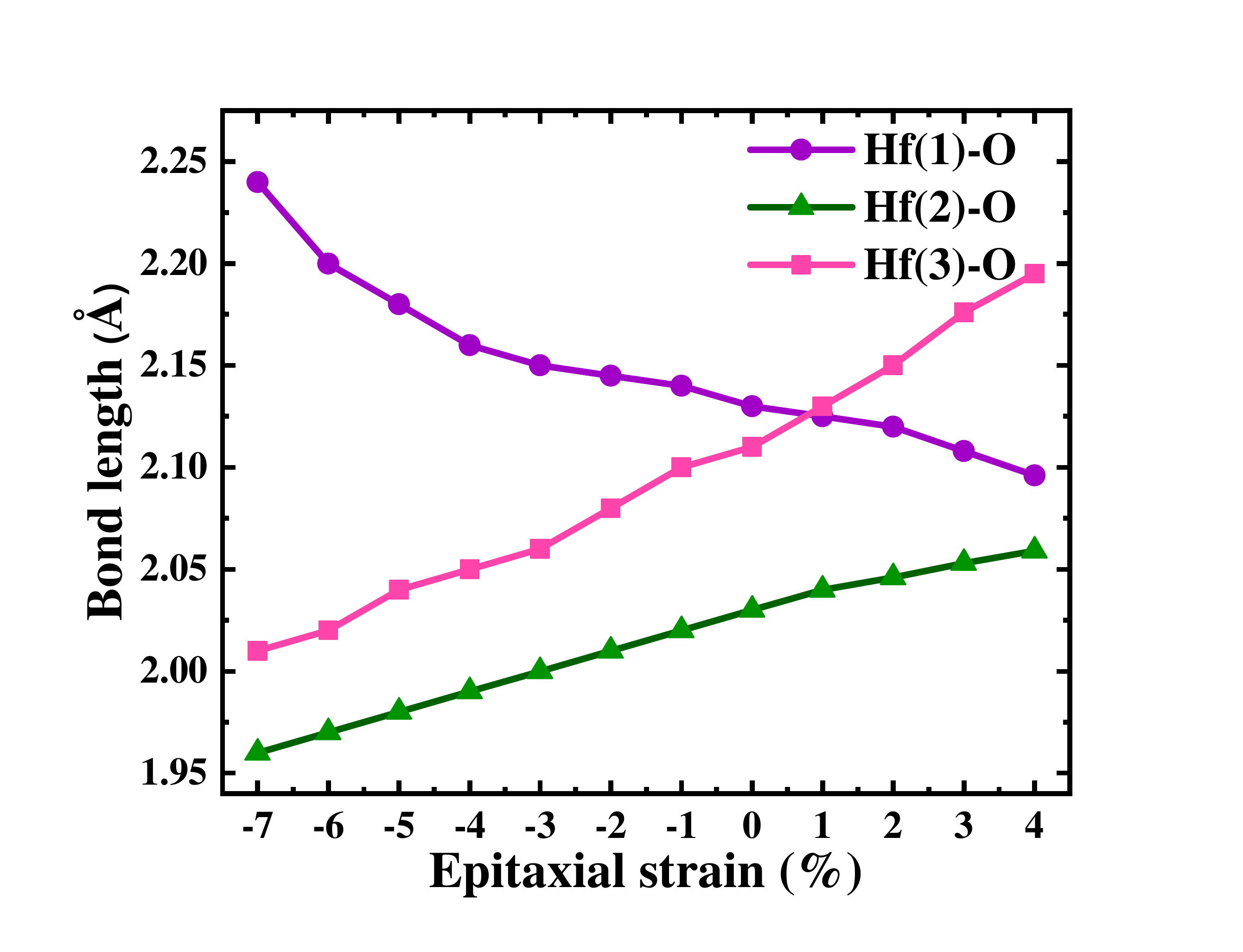}
    \caption{Lengths of the Hf(1)--O, Hf(2)--O and Hf(3)--O bonds defined in Figure~\ref{fig:pto}, computed as a function of epitaxial strain.}
    \label{fig:bonds}
\end{figure}

We do this by running structural relaxations where the in-plane
lattice vectors are constrained to form a 90$^{\circ}$ angle and their
magnitudes fixed to $a=a_{0}(1+\eta_{\rm epi})$ and
$b=b_{0}(1+\eta_{\rm epi})$, where $a_{0}$ and $b_{0}$ are the
previously obtained equilibrium lattice constants (Table~S1 in the
Supporting Information) and $\eta_{\rm epi}$ the applied epitaxial
strain. Our calculations suggest that the $Pca2_{1}$ orthorhombic phase is an
equilibrium energy minimum in a wide $\eta_{\rm epi}$-range, from about $-7$~\% to about $+4$~\%. (We find that beyond this range the ferroelectric polymorph losses its stability and transforms into other structures that are of no interest here.) Figure~\ref{fig:bonds} shows our results for the $\eta_{\rm epi}$ dependence of the bond lengths, which follow the
expected behavior: the Hf(1)--O link (which largely lies along the
vertical direction) gets longer as we compress in-plane ($\eta_{\rm
  epi} < 0$), following the growth of the out-of-plane lattice
constant $c$ (see Figure~S4 of the Supporting Information). Conversely, the Hf(2)--O and Hf(3)--O bonds (which are essentially parallel to the plane perpendicular to the polarization) shrink upon in-plane compression. We have thus achieved the desired control on the atomic environment of the O(I) atoms.

\begin{figure}[t!]
    \centering
    \includegraphics[width=0.5\textwidth]{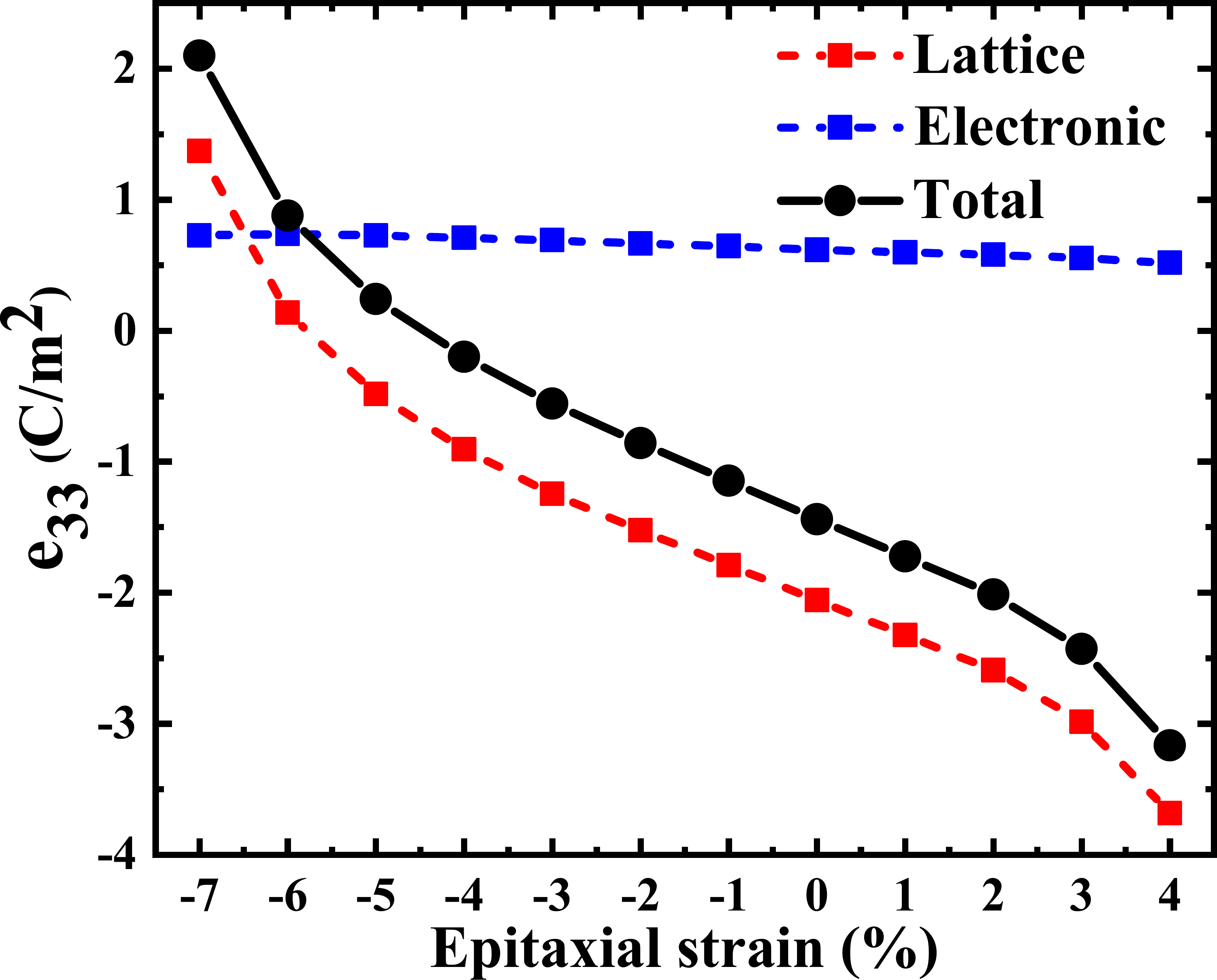}
    \caption{Computed $e_{33}$ piezoresponse component as a function of epitaxial strain. The total $e_{33}$ (black) is split into frozen-ion (blue) and lattice-mediated (red) contributions.}
    \label{fig:e33}
\end{figure}

Figure~\ref{fig:e33} shows the evolution of the $e_{33}$ piezoresponse component, as a function of $\eta_{\rm epi}$, obtained from DFPT calculations exactly as in the bulk case. We find that the frozen-ion
contribution $\bar{e}_{33}$ remains nearly constant (and positive) in
the whole range of strains. In contrast, the lattice-mediated part of
the response (red line in the Figure) changes very markedly in a
monotonic way. As a result, the total $e_{33}$ changes as well: it
reaches its strongest negative response at tensile strains ($\eta_{\rm
  epi}>0$) and eventually switches to positive values as we compress the material in-plane!
  
Let us stress that this change of sign in $e_{33}$ occurs even though we have a positive $P_{3}>0$ for all considered $\eta_{\rm epi}$ values. Indeed, as shown in  Figure~S5, we find that the polarization grows beyond 70~$\mu$C/cm$^{2}$ for epitaxial compression over $-$5~\%, an evolution that is perfectly consistent with that of the structural distortions (Hf--O bonds in Figure~\ref{fig:bonds}). Hence, the longitudinal piezoresponse changes sign even though the material remains in the same polar state; as far as we know, such an effect had never been observed (or predicted) before in a ferroelectric.

\begin{figure}[t!]
    \centering
    \includegraphics[width=0.5\textwidth]{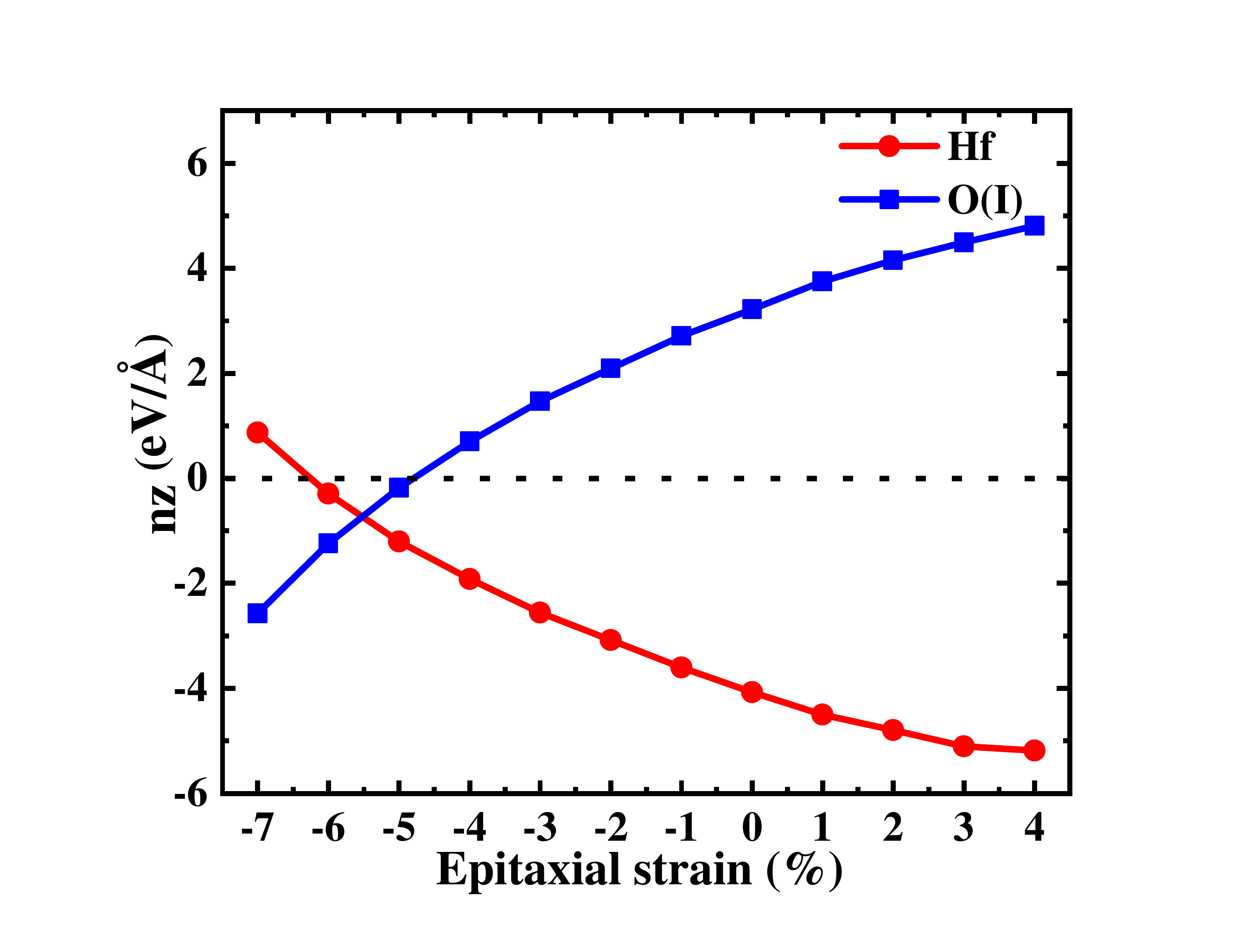}
    \caption{Epitaxial strain dependence of the $\boldsymbol{\Lambda}$ components that control the $e_{33}$ response (see text). More precisely, the shown components quantify the third (vertical) component of the force that acts on the Hf and O(I) atoms as a consequence of an applied strain $\eta_{3}>0$.} 
    \label{fig:lambda}
\end{figure}

We can easily track down the sign change of $e_{33}$ to the key
components of the force-response internal-strain tensor, whose
evolution with $\eta_{\rm epi}$ is shown in Figure~\ref{fig:lambda}. At zero strain, we have the situation already discussed above: when we
stretch ($\eta_{3}>0$) the state with $P_{3}>0$, the system's response
involves O(I) anions moving up (the corresponding $\Lambda_{nj}$
component is positive) and all Hf cations moving down (negative
$\Lambda_{nj}$ component), which results in a reduction of the
polarization $P_{3}$. Then, as we compress in-plane ($\eta_{\rm
  epi}<0$), the signs of the O(I) and Hf displacements eventually reverse, and so
does the piezoresponse to $\eta_{3}$.

\begin{figure}[t!]
    \centering
    \includegraphics[width=0.5\textwidth]{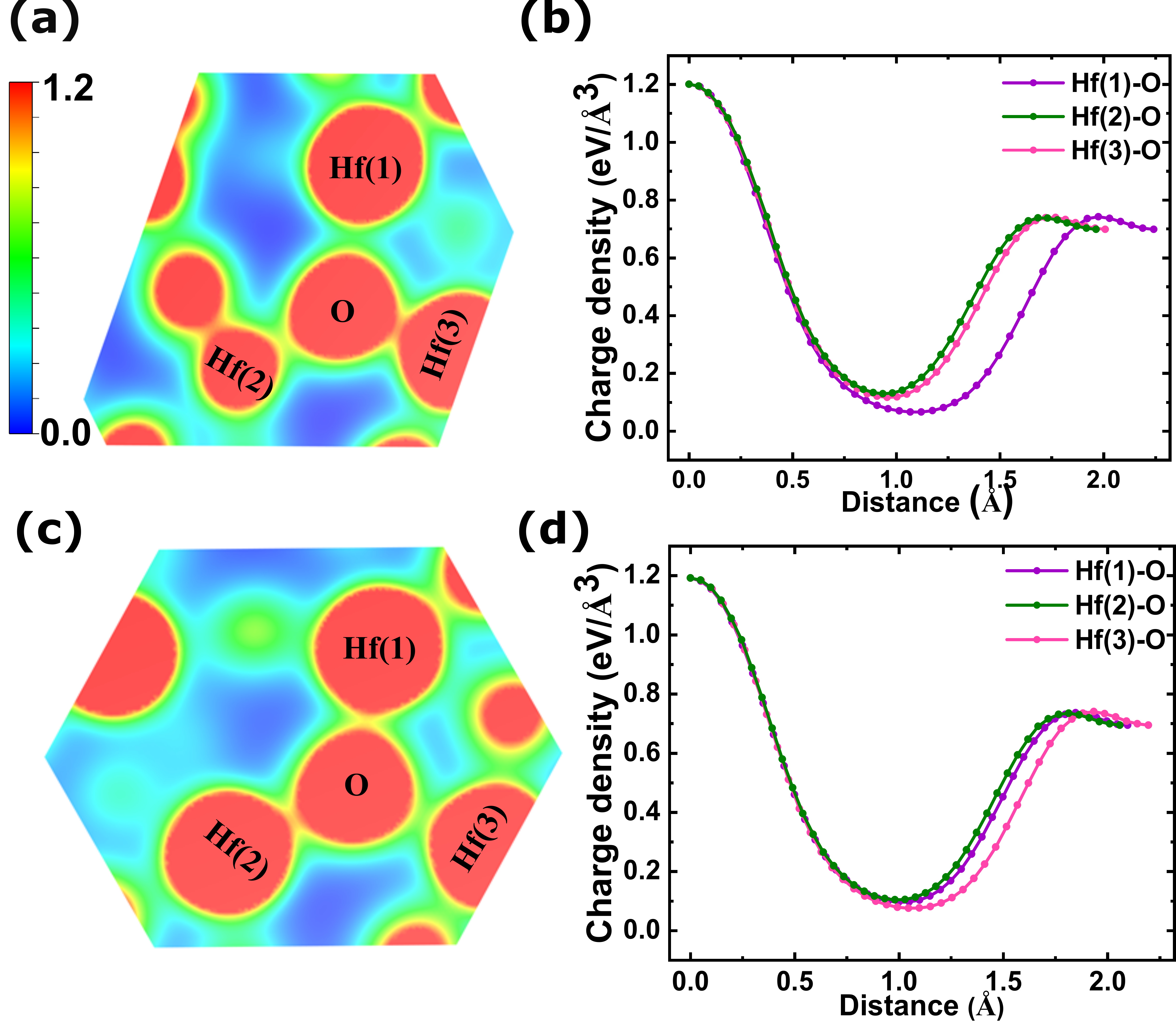}
    \caption{Same as Figure~\ref{fig:rho-eq}, but for the structures obtained at $\eta_{\rm epi} = -7$~\% ((a) and (b)) and $\eta_{\rm epi} = +4$~\% ((c) and (d)).}
    \label{fig:rho-epi}
\end{figure}

Finally, and most importantly, we can check whether our physical picture for the sign of $e_{33}$ is correct. The key results are given in Figure~\ref{fig:rho-epi}, which shows the computed charge density as obtained for the limit cases with $\eta_{\rm epi}=-7$~\%
(panels~(a) and (b)) and $\eta_{\rm epi}=+4$~\% (panels~(c) and (d)).

The result in Figure~\ref{fig:rho-epi}(a) and \ref{fig:rho-epi}(b) is particularly clear. For
strong in-plane compression the Hf(1)--O bond is all but broken, as
consistent with the long interatomic distance shown in
Figure~\ref{fig:bonds}. This suggests that, in this limit, the
piezoresponse of the material will be controlled by the Hf(2)--O and
Hf(3)--O links. Based on this assumption, we expect that a stretching
along the vertical direction will result in the O anion moving down
and the Hf cations moving up, so that these two bond lengths change as
little as possible. This is exactly what we find in our response
calculations; and this behavior results in the obtained $e_{33}>0$.

In the other limit ($\eta_{\rm epi}=+4$~\%,
Figure~\ref{fig:rho-epi}(c) and \ref{fig:rho-epi}(d)) we find that the Hf(1)--O and Hf(2)--O
bonds remain strong, while the Hf(3)--O seems relatively weak. Hence,
this case is similar to the bulk-like situation discussed above and corresponding to $\eta_{\rm epi}=0$~\%. The only difference is that the preponderance of the vertical Hf(1)--O bond can be expected to grow, which should result in a stronger $e_{33}<0$, as we indeed obtain.

Hence, our epitaxial-strain calculations confirm that our physical understanding of the longitudinal piezoresponse of HfO$_{2}$ is essentially correct: the peculiar atomic environment of the active oxygen atoms, and the tendency to maintain the optimal length of the dominant Hf--O bonds, determine the sign of $e_{33}$. Our calculations also show that the value of $e_{33}$ is strongly tunable, and can even change sign, provided one is able to act upon said atomic environment. Epitaxial strain gives us a control knob to do this.

\vspace{5mm}{\bf\cabin Summary and conclusions}

Our first-principles analysis reveal the atomistic reasons why the predicted longitudinal piezoelectric response of HfO$_{2}$ ($e_{33}$ or $d_{33}$) is negative. More specifically, we show that, when hafnia is strained along its polar axis, the material reacts by shifting the oxygen anions responsible for its spontaneous polarization, so as to best preserve the equilibrium distance of the corresponding Hf--O bonds. Naturally, this atomic rearrangement affects the polarization, in such a way that it grows when the strain is compressive, yielding a negative longitudinal effect. 

Guided by this observation, we are able to identify a strategy to tune the piezoresponse -- by controlling the chemical environment of the active oxygens --, showing that it can be enhanced or reduced, and even reversed to obtain a positive effect. Admittedly, the specific strategy tested here may not be applicable in practice. (We predict that large compressive epitaxial strains, beyond $-$5~\%, are needed to change the sign of $e_{33}$.) Nevertheless, our qualitative result is important: to the best of our knowledge, this is the first example of a ferroelectric whose piezoelectric response can be reversed by a continuous modification of the lattice, without switching its polarization. This possibility is unheard of among ferroelectrics, and certainly inconceivable in perovskite oxides. 

The theoretical prediction of a negative longitudinal piezoresponse is a robust one, corroborated in several ways by us and also obtained by other authors \cite{liu19b,liu20}. Further, we are not aware of any instance where the theoretical sign of the piezoresponse (as predicted by first-principles methods based on density functional theory, like the ones used here) contradicts the experimental observation. Hence, our experimental ratification of the negative effect -- by means of a careful piezoresponse force microscopy investigation of two reference ferroelectrics (PZT and PVDF) as well as HfO$_{2}$, all treated in exactly the same way so that a direct comparison can be made -- comes as no surprise. The reasonable quantitative agreement between the computed effect (about $-2.5$~pm/V) and the one estimated from experiments (between $-2$~pm/V and $-5$~pm/V) further strengthens our confidence in the results presented here. Note that the results of Ref.~\cite{chouprik19} on pristine Hf$_{0.5}$Zr$_{0.5}$O$_{2}$ samples also suggest a negative effect.

Having said this, it is important to recall that the vast majority of published experiments suggest a positive longitudinal piezoresponse $d_{33,{\rm eff}}$ \cite{muller11c,Starschich2014,schenk20}. Indeed, in the course of this work, we found ourselves that the application of the same experimental protocol to other HfO$_{2}$ films (thicker, grown by different means) yields a positive longitudinal effect. Hence, we find that different HfO$_{2}$ samples may present $d_{33,{\rm eff}}$ of different sign. This is a surprising observation, but one that resonates with our prediction that hafnia's piezoresponse can be reversed without switching its polarization. May the differences in the measured sign of $d_{33,{\rm eff}}$ be related to that?

It is thus clear that the experimental question of piezoelectricity in HfO$_{2}$-based compounds is still open and full of promise. Additional studies will be needed to evaluate how various factors (processing conditions, chemical composition, thickness, mechanical boundary conditions, electrical cycling) affect the outcome. We must try to correlate specific results for $d_{33,{\rm eff}}$ with specific (structural) features in the corresponding samples, distinguishing between intrinsic and extrinsic contributions to the response, a task for which first-principles theory may prove a valuable aid to experiment. This is a most appealing challenge, from both fundamental and applied perspectives. On the one hand, it may allow us to understand and master unprecedented ways to control piezoelectricity in ferroelectrics. On the other hand, it may allow us to optimize piezoelectricity in HfO$_{2}$ up to the point required for applications. We hope the present work will bring a new impetus to this effort.

\vspace{5mm}{\bf\cabin Methods.}

{\bf\cabin First-principles simulations.} Our calculations are carried out using first-principles density
functional theory (DFT) as implemented in the Vienna Ab-initio
Simulation Package (VASP) \cite{kresse96,kresse99}. We employ the
Perdew-Burke-Ernzerhof formulation for solids (PBEsol)
\cite{perdew08} of the generalized gradient approximation for the
exchange-correlation functional. In our calculations, the atomic cores
are treated within the projector-augmented wave approach
\cite{blochl94}, considering the following states explicitly: 5$d$,
6$s$, 6$p$ for Pb; 3$p$, 4$s$, 3$d$ for Ti; 2$s$, 2$p$ for O; and 5$s$, 5$p$, 6$s$, 5$d$ for Hf. To
calculate the response functions we use density functional
perturbation theory (DFPT) \cite{wu05}. All the calculations (for both
PbTiO$_{3}$ and HfO$_{2}$) are carried out using a plane-wave energy
cutoff of 600~eV. A 6$\times$6$\times$6 $k$-point sampling of the
Brillouin zone \cite{monkhorst76} is used for PbTiO$_{3}$
(corresponding to a 5-atom unit cell), while for HfO$_{2}$ we use
a 4$\times$4$\times$4 grid (corresponding to a 12-atom unit cell). The
structures are fully relaxed until the residual forces fall below
0.01~eV/Å and residual stresses fall below 0.1~GPa. We checked that
these calculation conditions yield well-converged results.

To verify our predictions for the piezoelectric properties of
HfO$_{2}$, we also run analogous DFPT calculations using the ABINIT
first-principles package \cite{gonze20}. In this case, we also consider the Perdew-Burke-Ernzerhof formulation of solids of the generalized gradient approximation for the exchange correlation functional. We use scalar relativistic norm-conserving Vanderbilt pseudo potentials as implemented in the ABINIT package \cite{Hamann2013}. In the calculations we treat explicitly the semicore states of Hf (5$s$, 5$p$, 4$f$, 5$d$ and 6$s$) and O (2$s$). We consider a plane wave cut-off energy of 60~hartree and a 4$\times$4$\times$4 k-point sampling of the Brillouin zone. We relax the structures until the residual forces fall below $10^{-6}$~hartree/bohr.

{\bf\cabin Sample preparation.} 200~nm-thick (111)-oriented PZT films with a Zr/Ti ratio of 40/60 were fabricated by magnetron sputtering on the Pt bottom electrode. Reactive ion etching was carried out to fabricate capacitors with 50~nm thick IrO$_2$ top electrodes with lateral dimensions of 80$\times$80~$\mu$m$^{2}$.

The 20~nm-thick La:HfO$_{2}$ films were grown by atomic layer deposition on TiN bottom electrodes and capped with a TiN top electrode. The TiN/La:HfO$_{2}$/TiN stack was then annealed in a N$_{2}$ atmosphere at 800$^{\circ}$C for 20~s. Details of the growth process for the La:HfO$_{2}$ films can be found in \cite{Schroeder2018}.

The 12~monolayer-thick (21.6~nm) PVDF films were deposited on Pt/Si substrates by Langmuir–Blodgett methods \cite{sharma11}.   

{\bf\cabin Sample characterization.} Switching spectroscopy PFM measurements were performed on a commercial atomic-force-microscopy system (MFP-3D, Asylum Research) in the resonance tracking mode using single crystalline diamond tips (D80, K-Tek, Nanotechnology) and Pt-coated tips (HQ:DPER-XSC11, MikroMasch). Electrical bias was applied to the top electrode using an external probe, with the frequency of the ac modulation signal around 350~kHz and 650~kHz for the D80 and the Pt-coated tips, respectively. For the PVDF thin films, the conducting tip acted as a local top electrode.  

In the DART mode, a feedback loop tracks a shift in the resonance frequency by measuring the difference in the PFM amplitudes for the two drive signals -- above and below the resonance frequency. The PFM loops shown in this work were all obtained below the resonance frequency.

\vspace{5mm}{\bf\cabin Data availability}

All the relevant data are available from the authors upon reasonable request.

\vspace{5mm}{\bf\cabin Code availability}

The first-principles simulations were done with VASP, which is
proprietary software for which the LIST group owns a license.

\vspace{5mm}{\bf\cabin Acknowledgements}

This work was funded by the Luxembourg National Research Fund (FNR) through grants PRIDE/15/10935404 “MASSENA” (S.D. and J.\'I.) and FNR/C18/MS/12705883 “REFOX” (H.A. and J.\'I.). Work at UNL was supported by the National Science Foundation through EPMD (Grant No. ECCS-1917635) Programs. Work at Namlab was financially supported out of the State budget approved by the delegates of the Saxon State Parliament.

\vspace{5mm}{\bf\cabin Author contributions}

S.D. performed the first-principles study, assisted by H.A. and
supervised by J.\'I. Samples were prepared by S.G., E.D, C.R. and
U.S., and the PFM characterization was carried out by P.B., H.L. and
A.G. The manuscript was written by J.\'I., S.D., A.G. and S.G., with
contributions from H.A., U.S. and E.D. J.\'I. conceived and
coordinated the work.

\end{document}